\documentclass[aps,prb,twocolumn,showpacs,floatfix]{revtex4-1}

\usepackage{graphicx}
\usepackage[encapsulated]{CJK}
\usepackage{color}
\usepackage{times}
\usepackage{ucs}

\begin{document}

\begin{CJK*}{UTF8}{} 

\title{Magnetism of epitaxial Tb films on W(110) studied by 
spin-polarized low-energy electron microscopy}

\author{J.E. Prieto}
\email{joseemilio.prieto@uam.es}
\affiliation{Centro de Microan\'alisis de Materiales, 
Dpto. de F\'\i{}sica de la Materia Condensada, IFIMAC and Instituto 
``Nicol\'as Cabrera,'' 
Universidad Aut\'onoma de Madrid, E-28049 Madrid, Spain}
\author{Gong Chen (\CJKfamily{gbsn}陈宫)}
\author{A.K. Schmid}
\affiliation{Lawrence Berkeley National Laboratory, Berkeley, California 94720, 
USA}
\author{J. de la Figuera}
\affiliation{Instituto de Qu\'{\i}mica F\'{\i}sica ``Rocasolano,'' CSIC, Madrid 
28006, Spain}

\date{\today}
\begin{abstract}
Thin epitaxial films of Tb metal were grown on a clean W(110) 
substrate in ultra-high vacuum and studied 
in-situ by low-energy electron microscopy. 
Annealed films present magnetic contrast in spin-polarized 
low-energy electron microscopy.
The energy dependence of the electron reflectivity was determined and
a maximum value of its spin asymmetry of about 1\% was measured.
The magnetization direction of the Tb films is in-plane. 
Upon raising the temperature, no change in the domain distribution is 
observed, while the asymmetry in the electron reflectivity 
decreases when approaching the critical temperature, following a power law 
$ \sim (1 - T/T_C)^\beta$ with a critical exponent $\beta$ of 0.39.

\end{abstract}


\maketitle

\end{CJK*}

\section{Introduction}

The hardest magnetic materials known to date~\cite{Rodewald_rare-earth_2007} 
are intermetallic systems containing magnetic rare earths (RE) alloyed with 
3d transition metals (TM), as in Co-Sm and Nd-Fe-B. In these compounds, 
the high magnetic anisotropies are induced by the RE ions, while the 
characteristic high ordering temperatures of the ferromagnetic TMs are 
retained~\cite{Mer93}. These magnetic materials are widely used in 
applications that require strong permanent magnetic fields, raising 
concerns about the availabity of the required rare earths. In consequence, 
lanthanides have been considered as model systems for the effect of 
increasing the anisotropy of TMs~\cite{Her91,Bus91}. 
Lanthanide metals themselves form a class of magnetic materials with 
rather different magnetic characteristics as compared to TMs. The 
negligible overlap between the partially filled electronic 4$f$ shells of
neighboring atoms in lanthanides leads to strongly localized magnetic 
moments, which in general contain both an orbital and a spin part. 
The localized character of the 4$f$ moments is also responsible for the
negligibly small direct exchange interaction between lanthanide ions. Instead,
they couple only indirectly through the valence-band electrons of the 
metal~\cite{Har74} (RKKY interaction), leading to ordering temperatures 
which are typically below room temperature (RT). The induced valence-band 
polarization gives only a minor contribution to the 
magnetization~\cite{Roe75}, in contrast to the predominantly
itinerant moments of ferromagnetic TMs.

In particular, heavy rare-earth metals are interesting magnetic materials 
due to their different magnetic properties despite their similar
crystalline and electronic structures. For example, Gd and Tb 
crystallize in the hexagonal close-packed structure with lattice
parameters that differ by less than 2\%. On the one hand, the spherical charge distribution of 
the half-filled 4$f$ shell of Gd leads to only a small magnetocrystalline anisotropy and 
hence to small
coercive fields in epitaxial films of good crystalline quality. On
the other hand, Tb shows a large magnetic anisotropy~\cite{Rhy67} due to its
non-spherical 4$f$-charge distribution caused by a large atomic orbital
momentum ($L$=3) which forces the magnetization to be in the basal plane. 
The easy axis is $\langle10\bar{1}0\rangle$ at all temperatures. 
Tb is ferromagnetic below 221~K, while between 221~K and 229~K it presents an helical 
magnetic structure than can be driven to a ferromagnetic arrangement by applied magnetic 
fields~\cite{Rhy72}. The magnetic moment per atom is 9.34~$\mu_{\rm{B}}$ with only a 
small contribution (of the order of the fractional part, 0.34~$\mu_{\rm{B}}$)  from itinerant 
electrons. 

Pure lanthanide-metal thin films have not been so widely studied as compared with TM films, 
in part due to their high chemical reactivity. For example just one study reported the 
distribution of magnetic domains in Dy films, by means of spin-polarized scanning 
tunneling microscopy~\cite{Ber07}. In particular, no results of low-energy electron
microscopy (LEEM) on lanthanide films have been reported to date. LEEM is a powerful 
technique allowing to visualize the surface morphology with a resolution of several 
nanometers and to study surface processes (e.g. crystal growth) in real time~\cite{Bau14}. 
If spin-polarized electrons are used as an illumination source 
(spin-polarized LEEM or SPLEEM~\cite{Rou10}), the magnetic domain distribution of the 
films can be imaged in real space by using exchange scattering. 
In this work, 
we prepared epitaxial films of Tb on W(110) and imaged their magnetic domains 
by means of SPLEEM.

\section{Experimental}

Epitaxial Tb metal films of thicknesses of up to about 20~ML were prepared 
in situ by vapor deposition in ultrahigh vacuum on a 
single-crystalline W(110) substrate, which had been cleaned previously by 
cycles of oxygen exposure and high-temperature flashing, following a 
procedure known to produce clean and ordered W(110) surfaces well suited 
for subsequent growth of metallic films~\cite{Sta00,Hei05,Rou06}.
A high-purity Tb rod heated by electron bombardment was used as 
evaporation source. Deposition rates were of the order of 0.1~nm per minute.
The base pressure in the chamber was in the 10$^{-11}$ Torr range
and rose to about 7$\times$10$^{-10}$ Torr when evaporating Tb.
The vacuum chamber was equipped with a conventional
electron optics for performing low-energy electron diffraction and a 
cilindrical mirror analyzer for recording Auger electron spectra.
The sample temperature was measured by means of a WRe thermocouple attached 
to a molybdenum plate on which the W(110) crystal rests.  
The absolute error of our temperature measurement arises from the lack of a 
cold reference, from additional junctions of different materials and from 
the contact point of the WRe thermocouple with a washer underneath the sample. 
It can be as high as 10-20~K, as determined by comparing selected 
transitions observed both with the WRe thermocouple and a Pt1000 resistor. 
However the relative error is much smaller, in particular in our
measurements as a function of the temperature, where no use was made of 
the sample heating filament so all the sample holder block was in thermal 
equilibrium with the cooling stage.

Spin-polarized LEEM measurements were performed in a low-energy electron 
microscope equipped with a spin-polarized electron source 
and a spin manipulator to adjust the spin direction of the electron beam 
with respect to the sample surface. Magnetic imaging is achieved in 
SPLEEM by representing the difference between LEEM images 
obtained with electron beams of opposite spin polarizations and
normalizing by the sum. 
The intensity in the resulting SPLEEM images depends on the projection of the
local surface magnetization onto the direction of spin polarization of the 
electron beam. As the electron-beam spin polarization can be 
changed with respect to the sample, the magnetization vector can be determined 
in real space with nanometer resolution~\cite{Ram04}. More details on both the 
instrument~\cite{Grz94}, the spin-polarization control method~\cite{Dud95}, 
or the vector magnetometric application of SPLEEM~\cite{Ram04,Gab06}
can be found in the literature. The electron beam energy 
is referred to zero energy which corresponds to the sample 
and the cathode at the same potential.

\section{Results and discussion}

\begin{figure}[t]
\includegraphics*[width=0.45\textwidth]{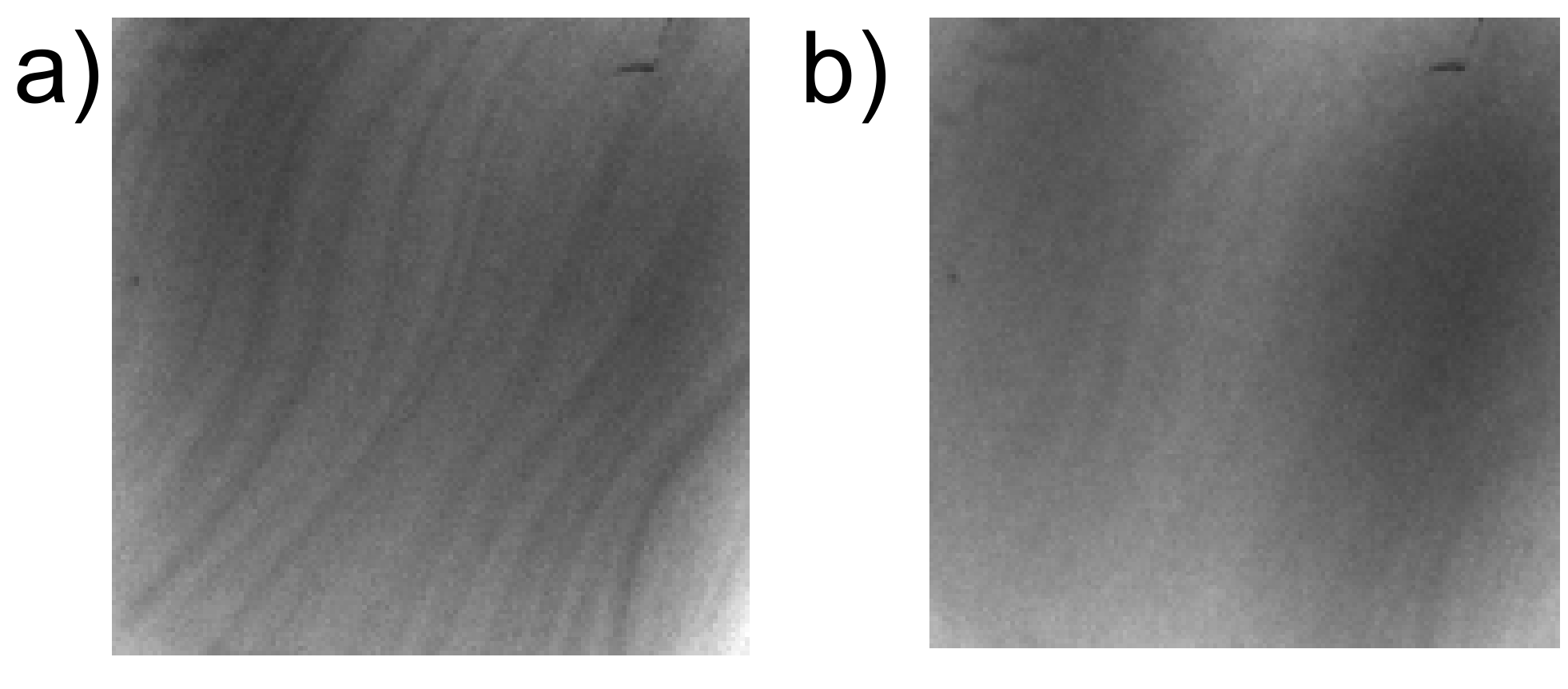}
\caption{\label{WandTbW} (a) LEEM image of the clean W(110) surface prior to
deposition.The electron energy
is 5.4~eV and the field of view is 7.5~$\mu$m.
(b) LEEM image of the as-grown 4~ML Tb/W(110) film.
The electron energy
is 6.2~eV and the field of view is 7.5~$\mu$m.}
\end{figure}

\begin{figure}[t]
\includegraphics*[width=0.45\textwidth]{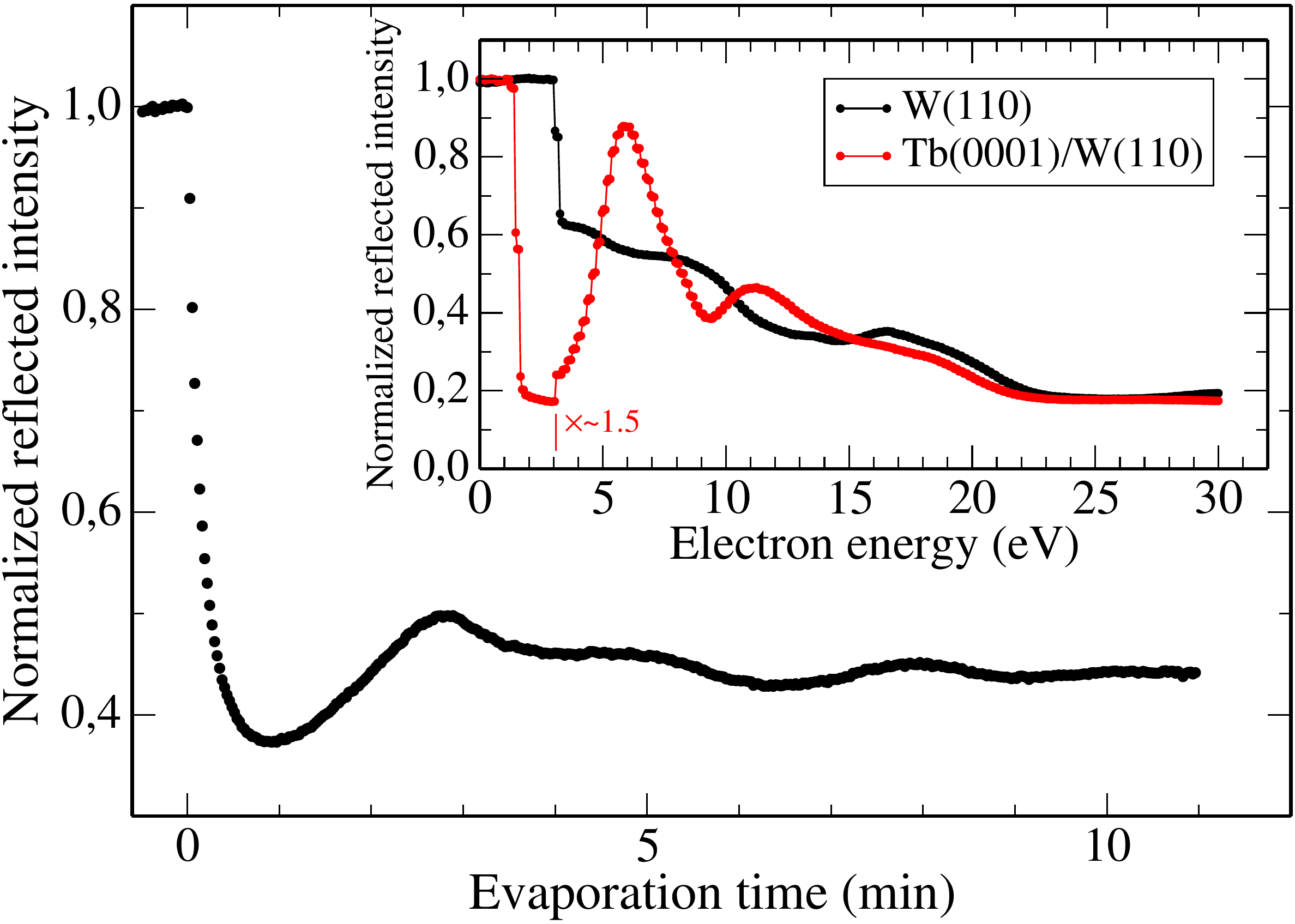}
\caption{\label{Tbgrowth} (color online) Electron reflectivity at an electron energy of 5.4~eV
as a function of the Tb evaporation time
on the clean W(110) surface, normalized to the value prior to deposition start.
Inset: Reflected electron intensity as a function of energy for both the 4~ML
Tb(0001)/W(110) film and the substrate W(110). Intensities have been
normalized to their respective maximum values and there is an increase in
gain by a factor of about 1.5 due to a change in the detector settings in the
curve for the film at an energy of about 3~eV.}
\end{figure}

A typical image of the clean W(110) surface prior to deposition shows 
step bunches, as can be seen in Fig.~\ref{WandTbW}a. 
Immediately after starting the deposition of Tb at room temperature, the step 
contrast is lost and a strong decrease of the reflected electron intensity is 
detected. In LEEM this is typically due to the nucleation of islands with 
sizes below the resolution limit of the microscope (typically $\sim$30~nm). 
Continuing the deposition, the reflected intensity reaches a minimum. 
After the minimum (at about 1 minute of evaporation time), 
the step bunches 
begin to be weakly detected again, and the averaged reflected intensity 
recovers partially and then oscillates with a small amplitude, as shown in 
Fig.~\ref{Tbgrowth}. 
Up to 4 maxima in the intensity can be seen for a deposition
time of about 11 min. A ratio of the intensities of the Auger electron
transitions Tb$_{146~{\rm eV}}$ and W$_{179~{\rm eV}}$ of 9.9 was 
determined experimentally. Using the values for the inelastic electron 
mean free paths from Seah and Dench~\cite{Sea79} and tabulated sensitivity 
factors~\cite{handbookaes}, 
a coverage of about 4.5~ML can be deduced, so that the observed intensity 
maxima can be attributed to the completion of successive monolayers.

The inset of Fig.~\ref{Tbgrowth} shows the reflected electron intensity as a 
function of the electron energy, both for the clean W(110) surface and for the 
4~monolayer (ML) Tb/W(110) film. Intensities have been normalized to their 
respective 
values at zero energy. The initial drop of the reflected intensity marks 
the transition from mirror mode,  where the electron energy lies 
below the work function of the surface, to regular diffraction 
imaging~\cite{Bau14}. The decrease in the transition energies  
indicates that the Tb film has a work function which is lower by 
1.7~eV than the bare W(110) surface, in fair agreement with the 
literature values of 4.55~eV for W and 3~eV for Tb~\cite{Dea99}.

\begin{figure}[t]
\includegraphics*[width=0.4\textwidth]{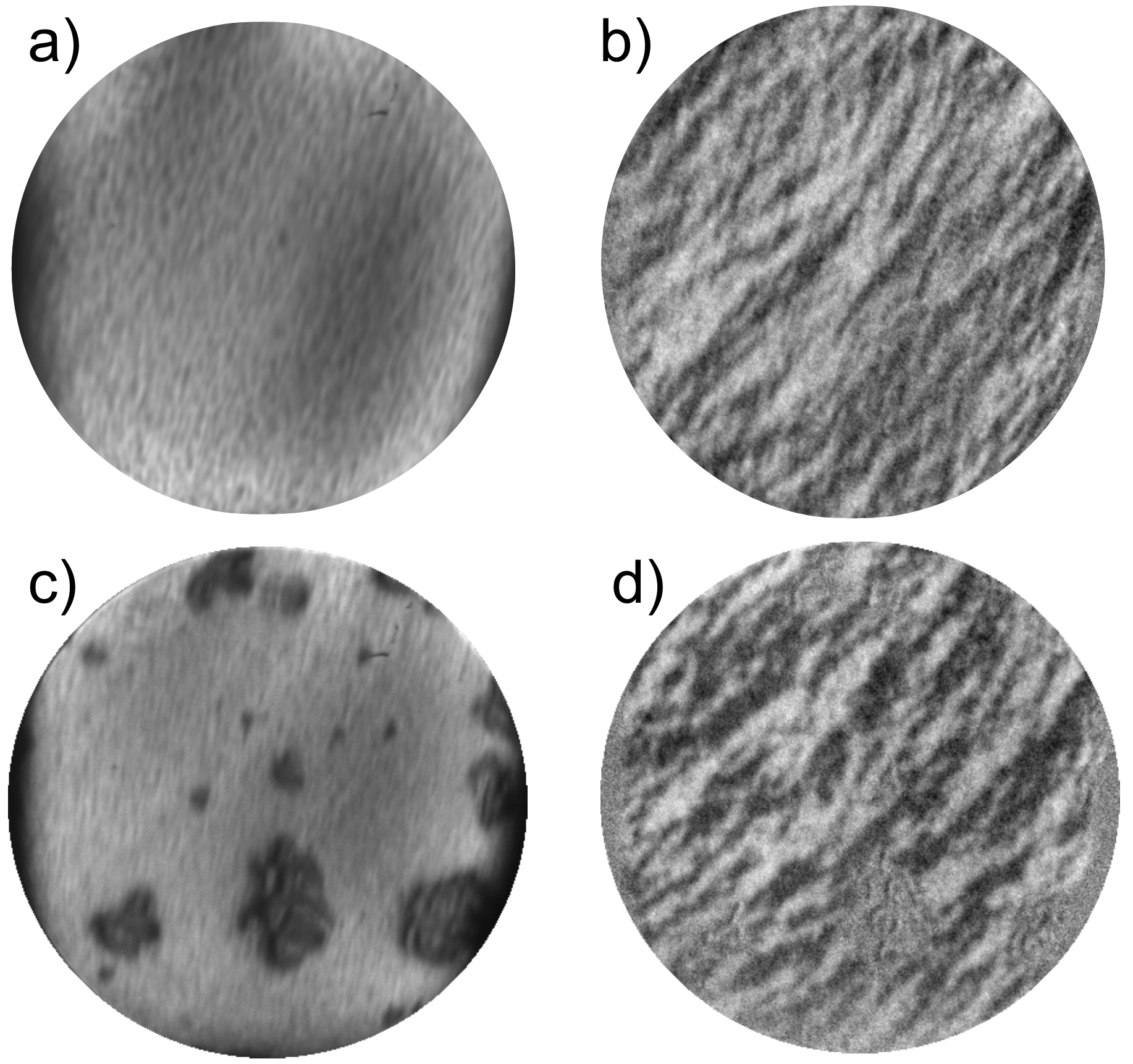}
\caption{\label{LEEM+SPLEEM} (a) LEEM image showing the film topography after
annealing at 650~K; (b) SPLEEM image with the electron polarization direction
in-plane with azimuthal angle of 70$^\circ$; (c) LEEM and (d) SPLEEM images
of a region containing holes due to dewetting of the substrate after
annealing at 800~K, showing magnetic contrast only in the film region.
The field of view is 12~$\mu$m.}
\end{figure}

An image of the as-grown Tb films is shown in Fig.~\ref{WandTbW}b.
The step structure of the substrate is barely visible. In order to obtain 
lanthanide films with well-defined magnetic properties, it is important 
to achieve a good epitaxial morphology by choosing the appropriate 
annealing temperatures~\cite{Asp94,Sta94}. 
Annealing the Tb films up to a temperature of 800~K leads to an increase 
of the reflected electron intensity (not shown) due to a smoothing of the film 
morphology. This is consistent with the significant reduction in the step 
density for annealing in the range of substrate temperatures of 600--800~K 
reported in Ref.~[\onlinecite{Hei05}]. 
In consequence, the Tb films were annealed up to a temperature of 650~K, 
and then cooled down to 80~K and analyzed by SPLEEM. 

Figure~\ref{LEEM+SPLEEM}a shows an annealed 20~ML Tb film at low temperature.
This film is expected to be thick enough to posess essentially 
bulk-like magnetic properties. For example, epitaxial films of 
Fe~\cite{Stm87} or Co~\cite{Sch90} of 5~ML thickness already show the Curie 
temperatures of the bulk materials. For certain 
electron energies and using an in-plane spin polarization direction of the 
electron beam, magnetic contrast is detected in the normalized 
pixel-by-pixel difference 
image of LEEM images acquired with opposite spin-polarizations, shown in 
Fig.~\ref{LEEM+SPLEEM}b. Black and white areas indicate regions where the 
magnetizacion has a non-zero component either parallel or antiparallel to 
the electron spin direction. The domain structure is ragged with domain 
walls that tend to follow the directions of the substrate steps, with 
typical widths of a few hundred nm. No component of the magnetization 
was detected in the out-of-plane direction. 

\begin{figure}[htb]
\includegraphics*[width=0.5\textwidth]{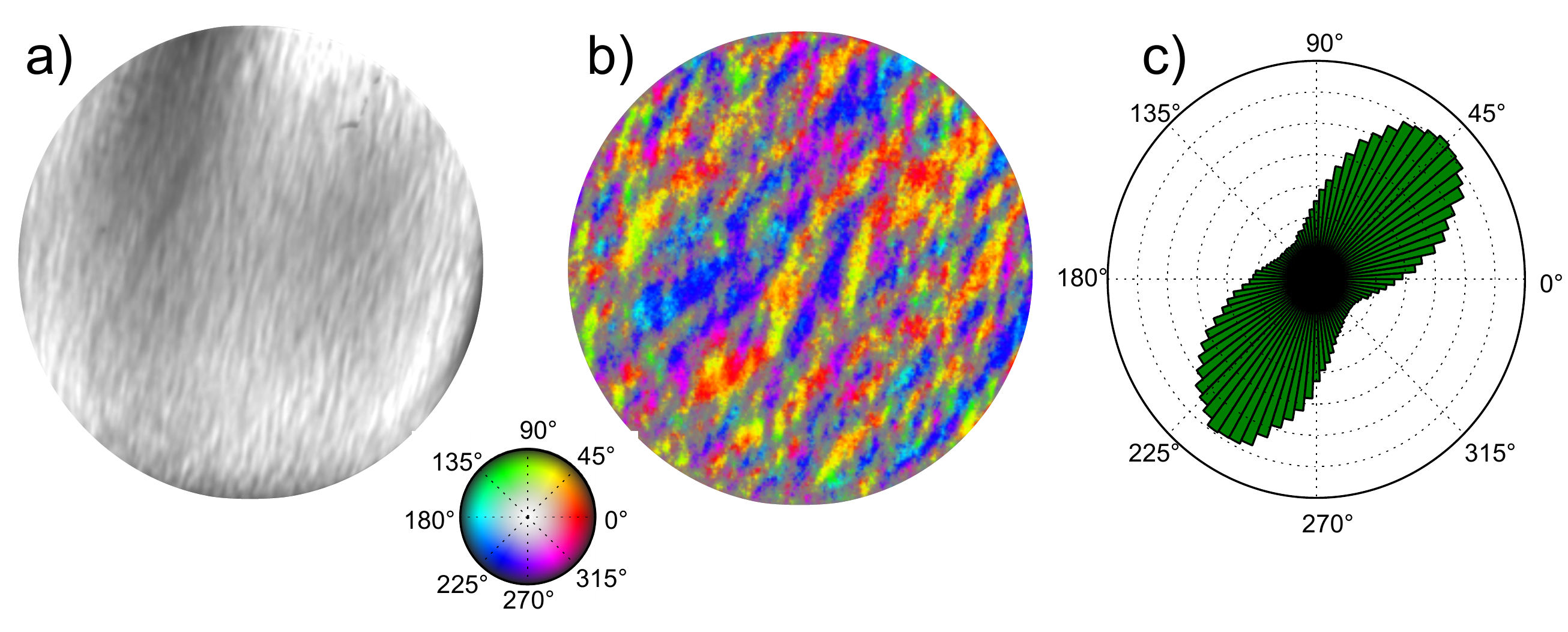}
\caption{\label{color} (color online) (a) LEEM image showing the film topography after
annealing at 650~K; (b) Composite color image combining SPLEEM images
acquired with the spin-polarization of the electron beam at 87 and -3 deg.
in-plane. (c) Polar histogram of magnetization directions in the area
imaged in (b). Field of view is 12~$\mu$m.}
\end{figure}

To determine the orientation of the magnetization on the Tb film, 
pairs of SPLEEM images were acquired with the electron beam spin polarization 
aligned in orthogonal directions. Pixel intensity in the individual 
images represents Cartesian components of the magnetization vector, 
so that the 2D magnetization vector can be mapped.
For the 20~ML Tb film shown in the LEEM image of 
Fig.~\ref{color}a, this has been represented in Fig.~\ref{color}b, 
where the magnetization orientation is given by the 
color according to the colormap shown below. To better visualize the angular 
distribution of the magnetization, a polar histogram~\cite{Mar16} is shown 
in Fig.~\ref{color}c. 
The histogram shown is based on an average of several pairs of images of 
the same region to improve the signal-to-noise ratio.
The observed magnetization follows a broad uniaxial distribution
with an axis of maximum contrast lying roughly at 45$\pm$20$^\circ$. 
This can be related to the crystallographic orientation of the sample. 
Tb/W(110) films grow in the Nishiyama-Wassermann epitaxial 
relation~\cite{Hei05}, i.e., with Tb[11$\overline{2}$0] parallel to W[001]. 
The LEED patterns of our Tb films (not shown) 
indicate that Tb[1$\overline{1}$00], which is parallel to 
W[1$\overline{1}$0], lies at 25$^\circ$ in the angular coordinates of
Fig.~\ref{color}. The observed preferred direction 
of spontaneous magnetization lies close to a Tb(0001) $b$-axis, the easy 
axis of bulk Tb metal~\cite{Rhy72}, in agreement 
with previous findings for epitaxial unmagnetized~\cite{Hei05} and 
remanently-magnetized Tb films~\cite{Dob07}. Nevertheless, other factors 
may introduce additional sources of anisotropy, such as the step distribution 
on the surface. The magnetic contrast in the images is produced by magnetic 
domains that extend preferentially along the main direction of the steps 
found in the substrate. Furthermore, the size of the domains visible in the 
images reaches several micrometers along the preferential step
orientation. 
This is in contrast 
with the high-resolution images recorded by spin-polarized scanning 
tunneling microscopy (SPSTM) of Dy films on W(110)~\cite{Ber07}, a system 
expected to be comparable to Tb/W(110). There, much smaller domains, 
with sizes between 40 and 800~nm, can be observed. Domains seen with SPLEEM 
in our work might therefore represent an average of smaller domains with 
different orientations.

\begin{figure}
\includegraphics[width=0.47\textwidth]{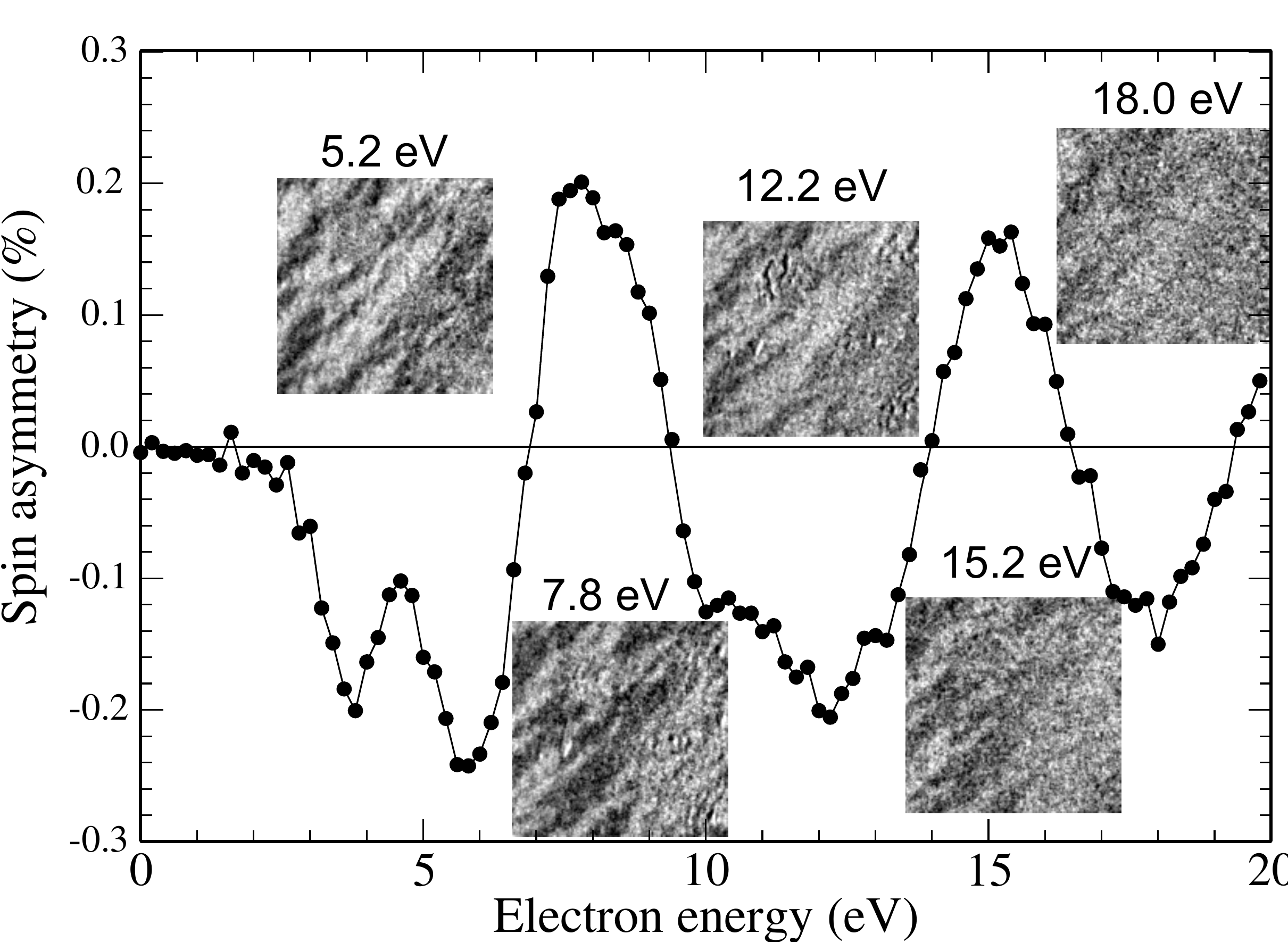}
\caption{\label{SPLEEMIV} Spin asymmetry of a 20~ML terbium film as a
function of electron energy along the 47$^\circ$ direction. The images
correspond to the energies indicated on top of each one and have a
size of 3.9~$\mu$m.}
\end{figure}

We have explored the dependence of the spin-asymmetry reflectivity on 
electron energy for a 20~ML Tb film. 
For this purpose, we measured SPLEEM images for different 
electron energies between 0 and 20~eV, and extracted the white-to-black 
contrast in the images. The results are shown in Fig.~\ref{SPLEEMIV}. 
The inserted images show an inverted contrast whenever the 
asymmetry curve changes sign (the apparent decrease of the magnetic 
contrast with increasing energy is due to the worsening of the 
signal-to-noise ratio as a result of the decreasing net reflectivity,
as shown by the inset of Fig.~\ref{Tbgrowth}.)  
The asymmetry was measured along the 
47$^\circ$ direction. The maximum of the asymmetry is around 0.2\%, which, 
for a degree of spin polarization of the electron beam of about 20\%, 
corresponds to a real asymmetry of the order of 1\%. 
We can compare this
value to the up to 10\% asymmetry measured in 3d-transition 
metals~\cite{Gra05}, or the 16\% asymmetry of magnetite~\cite{Fig13}. 
The experimental electron reflectivity has been compared in the past 
with theoretical calculations~\cite{Fle14}.
However, we are unaware of such calculations in spin-split systems.
The electronic states relevant for LEEM reflectivity are empty ones 
several eV's above the Fermi level. 
Spin asymmetry arises as a result of a difference in the density of states 
(DOS) for electrons of different spin orientations. This in turn depends 
on band splitting. 

Spin-dependent reflectivity of a ferromagnetic surface is a result
of two possible effects of the incoming spin-polarized beam scattering
on the electrons of the target in their spin-split electronic 
states~\cite{Rou10}. The first one is the spin dependence of the elastic 
scattering potential and the second is inelastic electron-electron 
collisions with the result of scattering into unoccupied states, whose
density of states (DOS) is different for both spin orientations.  
Since the inelastic mechanism leads to a reflectivity always higher for
minority electrons~\cite{Rou10}, the oscillatory character of the 
spin asymmetry shown in Fig.~\ref{SPLEEMIV} points to the presence
of elastic scattering. 
Furthermore, the reduced asymmetry detected in terbium films is probably
caused by the low spin polarization of the empty DOS induced 
by the localized $4f$ states of lanthanide metals, 
all of which appear at lower energies. In fact, experimental and 
theoretical results on the band structure of ferromagnetic Tb metal 
have shown an exchange splitting in the valence bands of about 
1~eV~\cite{Dob07}. 
This can be compared with the cases of Fe, with a band splitting of about
2~eV~\cite{San91} and magnetite with a difference of the order of 3~eV 
between corresponding states for spin-up and spin-down electrons~\cite{Fon07}.

\begin{figure}[ht]
\includegraphics*[width=0.45\textwidth]{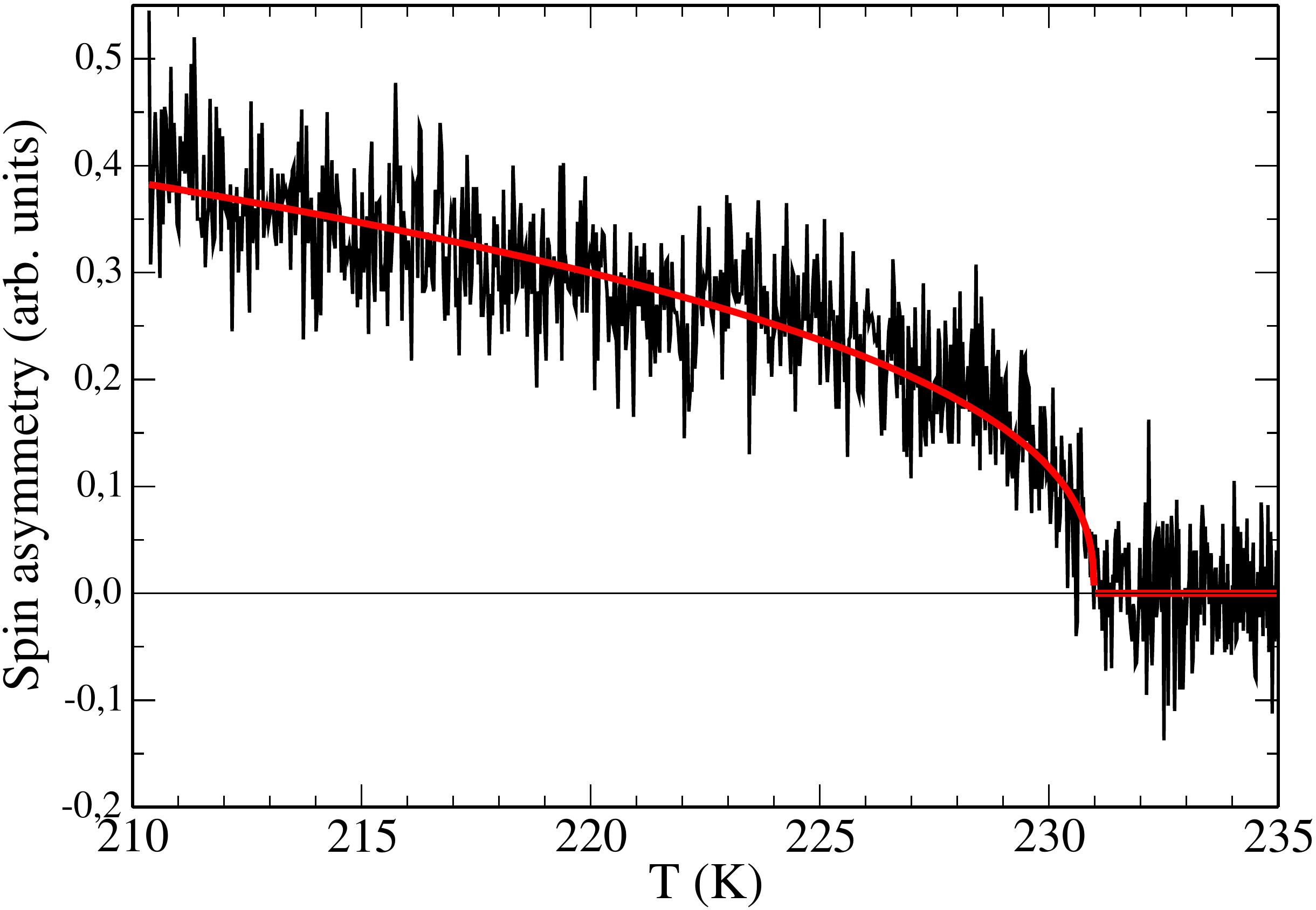}
\caption{\label{M-T} (color online) Plot of the spin asymmetry, proportional to the
sample magnetization, as a function of the temperature for a 20~ML Tb film
at an electron energy of 2.9~eV along the 47$^\circ$ direction.
The red line is a fit to an expression proportional to
$(T_C-T)^{\beta}$, giving $\beta$~=~0.39.}
\end{figure}

Annealing over 800~K leads to increased step bunching and 
eventually the film breaks up by opening pinholes 
(see Fig.~\ref{LEEM+SPLEEM}c) that extend down to the first Tb 
layer~\cite{Hei05}, as observed for other metal films such as Cr on 
W(110)~\cite{Mcc09}. Unlike Cr(110) films where the holes are strongly 
anisotropic, Tb(0001) pin-holes are more isotropic due to 
the hexagonal symmmetry of the film's atomic lattice. This behavior 
correlates with the onset of a decrease of the Curie temperature of 
the Tb/W(110) films revealed by measurements of magnetic susceptibility in
Ref.~[\onlinecite{Hei05}]. Figure~\ref{LEEM+SPLEEM}d shows that magnetic 
contrast is observed in the Tb smooth areas but not in the pinholes. 
The bottom of the holes has been proposed to be covered by a single Tb 
monolayer, as shown by the LEED pattern observed after annealing to these 
temperatures~\cite{Hei05} and in accordance with the behavior of different 
rare-earth films on W(110) at monolayer and submonolayer 
coverages~\cite{Kol86}. Our result shows that the ordering temperature of
this Tb monolayer lies below 80~K.

The evolution of the magnetic domains was followed upon raising 
the temperature and crossing the transition from the ferromagnetic 
to the helical antiferromagnetic phase. The distribution and shape 
of the domains does not change at all with temperature, suggesting 
that the domains are pinned down to structural defects. In fact, 
magnetic domains have been observed to be closely linked to the 
film morphology in the Dy/W system~\cite{Ber07}. However, the 
magnetic contrast itself, measured by the spin-asymmetry in the 
reflectivity at a constant electron energy, changes with temperature. 
The magnetic contrast decreases upon increasing the temperature, 
disappearing at a temperature of 231~K, as shown in Fig.~\ref{M-T} 
for a 20~ML Tb/W(110) film. The magnetic contrast in SPLEEM can be 
considered a proxy of the magnetization. Thus the plot in 
Fig.~\ref{M-T} corresponds to the evolution of the magnetization 
with temperature. We note that the transition from the 
ferromagnetic to the helical antiferromagnetic phase is a first-order 
transition, while the transition from helical to paramagnetic is
a conventional second-order magnetic-ordering transition.
While SPLEEM is highly surface sensitive, for electron energies only
a few eV's above the Fermi level, it actually probes several atomic 
layers~\cite{Bau14}, so the helical antiferromagnetic order is not 
expected to provide a significant contrast in SPLEEM. 
We thus assume that the disappearance of the magnetic contrast at 
the measured temperature of 231~K corresponds to the Curie 
temperature $T_C$ of the film.  
This is in reasonable agreement with measurements by magnetic 
susceptibility on annealed Tb films~\cite{Hei05}, 
where a similar value of the $T_C$ of films annealed to 
these temperatures is observed. 

A fit to an expression $M \sim (T_C-T)^{\beta}$ in Fig.~\ref{M-T} 
gives an exponent $\beta$~=~0.39$\pm$0.02. This value is in 
remarkable coincidence with those found for other lanthanide metals, 
including systems with low (Gd, Ref[\onlinecite{Sal82}]) 
and high (Ho, Ref[\onlinecite{Eck76}]) values of the 
magnetocrystalline anisotropy. This confirms the
bulk-like critical behaviour of our 20~ML thick films.
Compared with the values of the critical exponent 
$\beta$ for the three-dimensional (3D) models of Ising (0.326), 
X-Y (0.34), Heisenberg (0.365) and mean field (0.5), 
the value of 0.39 found in lanthanides has been
interpreted for Gd in favour of a basically Heiseberg-like critical 
behaviour with dipolar contributions~\cite{Fre97}. 

\section{Summary}

The magnetic domain structure of terbium films grown on W(110) was 
observed by SPLEEM. 
The polarization of the unoccupied density of states by the localized 
4$f$ levels gives rise to a spin-asymmetry reflectivity that was
measured as a function of electron energy and reaches a maximum value 
of the order of 1\%.
The local orientation of the magnetization was detected by 
combining SPLEEM information along different angles. The ferromagnetic 
to helical antiferromagnetic transition was followed in real space
while raising the temperature. While the domain distribution does not 
change through the transition, the magnetic contrast and thus the 
magnetization follows a critical exponent of 0.39, similar to the 
critical exponent measured by averaging techniques in different 
lanthanide metals such as Gd and Ho. 

\acknowledgments This research was partly supported by Spain under Projects No. 
MAT2014-52477-C5-5-P, MAT2015-64110-C02-1-P (MINECO) and 
FIS2008-01431 (MICINN). Experiments were 
performed at the National Center for Electron Microscopy, Lawrence Berkeley 
National Laboratory, supported by the Office of Science, Office of Basic Energy 
Sciences, Scientific User Facilities Division, of the U.S. Department of Energy 
under Contract No. DE-AC02-05CH11231.

\bibliography{terbium}


\end{document}